\documentclass[aps,prl,
showpacs,superscriptaddress,groupedaddress]{revtex4}  
\usepackage{graphicx}  
\usepackage{dcolumn}   
\usepackage{bm}        
\usepackage{amssymb}   

\begin{document}



\title{General structure of two-photon S matrix in waveguide quantum electrodynamics systems containing a local quantum system with multiple ground states}
\author{Shanshan Xu}
\email{xuss@stanford.edu}
\affiliation{Department of Physics, Stanford University, Stanford, California 94305}

\author{Shanhui Fan}
\email{shanhui@stanford.edu} \affiliation{Department of Electrical Engineering, Ginzton Laboratory, Stanford University, Stanford, California 94305}

\date{\today}

\begin{abstract}
We present the general structure of  two-photon S matrix for a waveguide coupled to a local quantum system that supports multiple ground states.
The presence of the multiple ground states results in a non-commutative aspect of the system with respect to the exchange of the orders of photons.
Consequently, the two-photon S matrix significantly differs from the standard form as described by the cluster decomposition principle in the quantum field theory.

\end{abstract}

\maketitle


The scattering matrices (S matrices) are of essential importance for characterizing the interaction of quantum particles. On one hand, each element of a scattering matrix describes the probability amplitude of a particular scattering event. 
Thus every element of a scattering matrix is of direct experimental significance. On the other hand, from a theoretical point of view, the analytic structure of an S matrix is strongly constrained by symmetries and causalities,
as well as by other general aspects such as the local nature of the interactions. Consequently, much of the literature on quantum field theory is devoted to the computation and elucidation of the structure of S matrices \cite{sw, qft1, qft2, qft3}. 
Using the cluster decomposition principle
\cite{ wc, taylor,sw}, the standard form of two-particle S matrix listed in quantum field theory textbooks is $S=S^0+i\,T$,
where $S^0$, the non-interacting part of the S matrix, is of the form
\begin{eqnarray}\label{oS0}
S^0_{p_1p_2k_1k_2}=t_{k_1}t_{k_2}\left[\delta(p_1-k_1)\delta(p_2-k_2)+\delta(p_1-k_2)\delta(p_2-k_1)\right]
\end{eqnarray}
and contains the product of two $\delta$ functions. The T matrix, which describes the interaction, is of the form
\begin{eqnarray}\label{oT}
T_{p_1p_2k_1k_2}=C_{p_1p_2k_1k_2}\delta(p_1+p_2-k_1-k_2)
\end{eqnarray}
and contains a single $\delta$ functions. Here, $k_{1,2}$ and $p_{1,2}$ are the momenta of the incident and outgoing particles, respectively. $t_k$ is the individual particle transmission amplitude and $C_{p_1p_2k_1k_2}$ characterizes 
the strength of the interactions between 
two particles.  Recently, this form is also shown to apply in waveguide quantum electrodynamics (QED) systems, where a few waveguide photons interact with a local quantum system 
 \cite{sf,sfA,fks,lsb,eks,zb,r,koz,lhl,sek,rwf,smzg,ll,sfs,rf,jg,lnsa,srf}. 
 
 In this letter, we show that there in fact exists a class of waveguide QED systems, in which the two-photon S matrix does not have the form of (\ref{oS0}). The key attribute of these systems is that the local quantum system has
 multiple ground states. We show that this attribute results in a non-commutative aspect of the system with respect to the exchange of the orders of photons, which strongly constrains the form of the S matrix. 
 This is in contrast to a large number of systems previously considered that have S matrix of the form shown in (\ref{oS0}). In these systems the local quantum system has a unique ground state and hence does not have
 such non-commutative property. 
 
 The results  here point to a much richer set of analytic properties in the structure of S matrix than previously anticipated. Also, examples of local quantum system with multiple ground states include
three-level $\Lambda$-type  atomic systems, which support two ground states in the electronic levels, as well as optomechanical cavities where the lowest lying photon-state manifolds contain multiple phonon sidebands. 
The three-level $\Lambda$-type systems play an essential role in constructing quantum memory and quantum gates for photons \cite{dk, csdl, kin, zgb2}, 
whereas reaching the photon-blockade regime with optomechanical cavities has been a long-standing experimental objective in quantum optomechanics \cite{rabl, akm}. 
Exploring the nature of photon-photon interaction in these systems in the context of waveguide QED is therefore of significance in a number of directions that are of importance for quantum optics. 
While there have been several calculations on the two-photon scattering properties of these systems \cite{roy, pg, zgb, ll2}, 
there have not been any discussions on the general analytic structure of the two-photon S matrix in this class of systems.  

\begin{figure}[h]
\includegraphics
[width=0.9\textwidth] {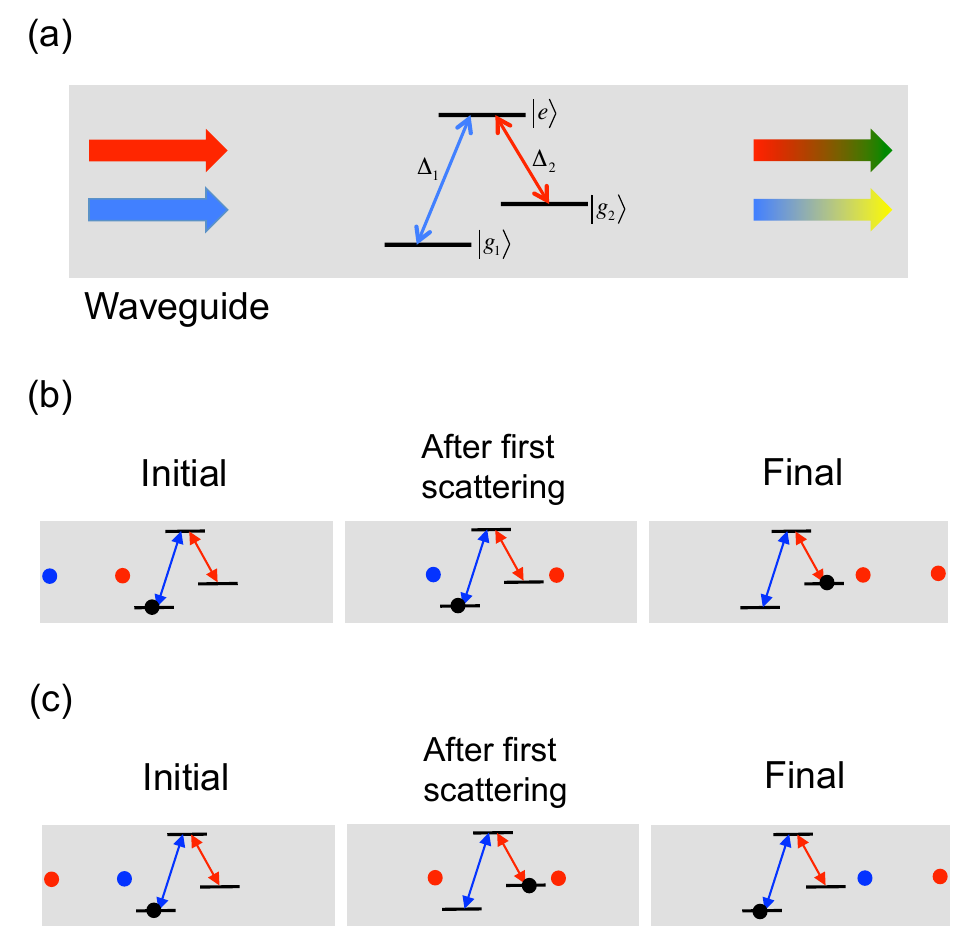} \caption{(a) The system we consider: a photonic waveguide
coupled to a three-level $\Lambda$-type atom. 
(b)  A sequential scattering event where two photons incident from the left scatter against a three-level atom with  $\gamma_1=\gamma_2\ll \Delta_1, \Delta_2$. 
The photons are represented by red or blue colors with different colors representing different frequencies of photons.
 We send in the red photon followed by the blue photon, in which the initial, intermediate and final
states are shown in the subplots. (c) Another sequential scattering event that is the same as (b) except for the reverse photon ordering.
} \label{fig1}
\end{figure}

We start by considering the simplest example of a single-mode waveguide coupled to a three-level $\Lambda$-type atom as shown in Fig.\ref{fig1} (a).
The Hamiltonian is described as
\begin{eqnarray}\label{int}
H&=&\int
dk\,k\,c_k^{\dag}\,c_k+ \sum_{\lambda=1}^2 \widetilde{\Delta}_{\lambda} |g_{\lambda}\rangle\langle g_{\lambda}|+\Omega |e\rangle\langle e|+\sum_{\lambda=1}^2\sqrt{\frac{\gamma_{\lambda}}{2\pi}}
\int dk \left(c_k^{\dag}\,|g_{\lambda}\rangle\langle e|+|e\rangle\langle g_{\lambda}|c_k\right)\,,
\end{eqnarray}
where $c_k\, (c_k^{\dag})$ is the annihilation (creation) operator
of the photon state in the waveguide. These operators satisfy the
standard commutation relation $[c_k, c_{k'}^{\dag}]=\delta(k-k')$.
 Here for simplicity we consider
a waveguide consisting of only a single mode in the sense of
Ref.\cite{sfA}. The argument here, however, can be straightforwardly
generalized to waveguides supporting multiple modes.
$\widetilde{\Delta}_1$, $\widetilde{\Delta}_2$ and $\Omega$ are the respective energy of the ground states $|g_1\rangle$, $|g_2\rangle$ and the excite state $|e\rangle$ of the atom satisfying
$\widetilde{\Delta}_1 <\widetilde{\Delta}_2<\Omega$.  
We define ${\Delta}_{\mu}\equiv \Omega-\widetilde{\Delta}_{\mu}$ for $\mu=1,2$.
The waveguide photons couple to both
$|g_1\rangle -|e\rangle$ and $|g_2\rangle -|e\rangle$ transitions of the atom with respective coupling constants $\sqrt{\gamma_1/2\pi}$ and $\sqrt{\gamma_2/2\pi}$. 
In general we assume that $\gamma_1, \gamma_2 \ll \Delta_1, \Delta_2$.
The single-photon S matrix for this system is 
\begin{equation}\label{singleS}
\left[\mathbf{S}_{pk}\right]_{\mu\nu}\equiv \langle p, g_{\mu}|S|k,g_{\nu}\rangle = t_{\mu\nu}(k)\, \delta(p-\Delta_{\mu}-k+\Delta_{\nu})\,,\label{singleS}
\end{equation}
where $\mu,\nu$ take values of $1,2$ and
\begin{equation}
t_{\mu\nu}(k)=\delta_{\mu\nu}-i\frac{\sqrt{\gamma_{\mu}\gamma_{\nu}}}{k-\Delta_{\nu}+i\left(\frac{\gamma_1}{2}+\frac{\gamma_2}{2}\right)}\label{tuv}
\end{equation}
is the transmission amplitude of the waveguide photon $|k\rangle$ when the initial and final states of the atom
are $|g_{\nu}\rangle$ and  $|g_{\mu}\rangle$, respectively \cite{tl,ws,klyn,fb}.

We proceed to provide an intuitive argument about the structure of the two-photon S matrix. As an example,
we consider a specific three-level system where $\gamma_1=\gamma_2$. For notation simplicity, we refer photons with energy $\Delta_1$ and $\Delta_2$ as  "blue" and "red" photons, respectively.
From (\ref{singleS}) and (\ref{tuv}), if the atom is initially in the ground state $|g_1\rangle$, an incident blue photon will be on resonance to the atomic transition. Therefore, upon scattering against the atom, it will be converted to a red photon while 
the atomic state is changed to $|g_2\rangle$,
whereas an incident red photon in the same situation will pass through the atom unchanged without affecting the atomic state, since it is off resonance from the atomic transition. 
A complementary behavior occurs when the atom is initially in the ground state $|g_2\rangle$, as can be deduced
from (\ref{singleS}) and (\ref{tuv}).

To illustrate the structure of the non-interacting part of the S matrix, we now construct a thought experiment as shown in Fig.\ref{fig1} (b) and (c) by considering the outcome of two different sequential scattering events where two photons are sent toward the atom
with a sufficiently large time delay between the two photons. In both events, we assume that the atom is initially in the 
ground state $|g_1\rangle$. In the first event (Fig.\ref{fig1} (b)),  we send in the red photon first, it passes by the atom without interaction. The blue photon then comes in
 and scatters against the atom. The scattering changes the atomic state from $|g_1\rangle$ to  $|g_2\rangle$, with the photon converted to red. Therefore, at the end of the two-photon scattering event, we end up with two red photons
and the atom in the state $|g_2\rangle$. In the second event (Fig.\ref{fig1} (c)),  we send in the blue photon first and then the red photon.
With a similar analysis as discussed above, we can show that we will end up with the red photon first and then the blue photon, with the atomic state remaining in $|g_1\rangle$.
In this system, the outcome of a two-photon scattering event depends on the order of the photons being sent in. We note that each of two different incident states above can be described by a symmetrized two-photon wavefunction. The two states are mapped to each other, not by an exchange symmetry operator, but rather by an operator 
$\hat{R}$ that exchanges the order of the photons. The observation above then indicates that $\left[\hat{R}, S\right] \neq 0$. Such a non-commutivity with respect to photon-order exchange operator arises from the existence of multiple ground states in the local quantum system. For local quantum system with a unique ground state, one can easily show with a similar thought experiment \cite{srf} that the outcome of the two-photon sequential scattering does not depend on the orders of the photons sent in.

The non-commutivity between the two-photon S matrix and photon-order exchange operator points to interesting aspects of the structure of two-photon S matrix. 
The two-photon S-matrix is typically computed with respect to a two-photon symmetrized plane wave:
\begin{eqnarray}
\psi_{\text{in}}(x_1, x_2)\equiv \frac{1}{2\sqrt{2}\pi}\left(e^{ik_1x_1}e^{ik_2x_2}+e^{ik_1x_2}e^{ik_2x_1}\right)\,.\label{iw}
\end{eqnarray}
To apply the argument above, we decompose $\psi_{\text{in}}(x_1, x_2)=\psi^{(1)}_{\text{in}}(x_1, x_2)+\psi^{(2)}_{\text{in}}(x_1, x_2)$, where 
\begin{eqnarray}
\psi_{\text{in}}^{(1)}(x_1,x_2)&=& \frac{1}{2\sqrt{2}\pi}\left[e^{ik_1x_1} e^{ik_2x_2}\,\theta(x_1-x_2)+e^{ik_1x_2} e^{ik_2x_1}\,\theta(x_2-x_1)\right]\,,\label{in1}\\
\psi_{\text{in}}^{(2)}(x_1,x_2)&=&\ \frac{1}{2\sqrt{2}\pi}\left[e^{ik_1x_1} e^{ik_2x_2}\,\theta(x_2-x_1)+e^{ik_1x_2} e^{ik_2x_1}\,\theta(x_1-x_2)\right]\,.\label{in2}
\end{eqnarray}
With the $\theta$ functions in (\ref{in1}) and (\ref{in2}), 
$\psi_{\text{in}}^{(1)}(x_1,x_2)$ can be viewed as the plane wave limit of two sequential single-photon pulses with the center frequencies of the leading and the trailing pulses centering at $k_1$ and $k_2$, respectively, while 
$\psi_{\text{in}}^{(2)}(x_1,x_2)$ is the limit of the same two pulses but with the order of the center frequency reversed.  
We now consider all the scattering pathways in which the atom changes from state $|g_{\nu}\rangle$ to $|g_{\mu}\rangle$ through the two-photon sequential scattering process.   
 For $\psi^{(1)}_{\text{in}}(x_1, x_2)$, the photon with frequency $k_1$ arrives first. 
 As one of the many possible scattering pathways, 
 upon scattering of this photon, the atom is driven from the state $|g_{\nu}\rangle$ to a ground state $|g_{\lambda}\rangle$, 
 whereas the wavefunction of the outgoing photon takes the form of
$\phi_{k_1\lambda\nu}(x_1)\equiv t_{\lambda\nu}(k_1)e^{i(k_1-\Delta_{\nu}+\Delta_{\lambda})x_1}/\sqrt{2\pi}$. Then the photon with frequency $k_2$ arrives. It drives the atom from the state $|g_{\lambda}\rangle$ to the state  $|g_{\mu}\rangle$, 
and as a result is converted to an outgoing photon with the wavefunction $\phi_{k_2,\mu\lambda}(x_2)\equiv t_{\mu\lambda}(k_2)e^{i(k_2-\Delta_{\lambda}+\Delta_{\mu})x_2}/\sqrt{2\pi}$. 
Summing over all the pathways as labelled by $\lambda$, the final state associated with $\psi^{(1)}_{\text{in}}(x_1, x_2)$ is then
$\psi^{(1)}_{\text{out}}(x_1, x_2)=\frac{1}{\sqrt{2}}\sum_{\lambda}\phi_{k_2,\mu\lambda}(x_2)\phi_{k_1,\lambda\nu}(x_1)\theta(x_1-x_2)+\left[x_1 \longleftrightarrow x_2\right]$. 
Consider both $\psi^{(1)}_{\text{in}}(x_1, x_2)$ and $\psi^{(2)}_{\text{in}}(x_1, x_2)$, the sequential scattering process then leads to the final state
\begin{eqnarray}
\psi_{\text{out}}(x_1,x_2)&=&\psi^{(1)}_{\text{out}}(x_1,x_2)+\psi^{(2)}_{\text{out}}(x_1,x_2)\nonumber\\
&=&\frac{1}{2\sqrt{2}\pi}\sum_{\lambda=1}^2 t_{\mu\lambda}(k_2) t_{\lambda\nu}(k_1)e^{i(k_2-\Delta_{\lambda}+\Delta_{\mu})x_2}e^{i(k_1-\Delta_{\nu}+\Delta_{\lambda})x_1}\theta(x_1-x_2)+\left[x_1 \longleftrightarrow x_2, k_1 \longleftrightarrow k_2\right]\,.\label{ow}
\end{eqnarray}
We note that the $\theta$ functions in (\ref{ow}) don't compensate each other, as a direct result of the non-commutivity in the sequential scattering process. From (\ref{ow}), by Fourier transformation, we obtain the 
the non-interacting part of the two-photon S matrix as
\begin{eqnarray}\label{seq}
\left[\mathbf{S}^{0}_{p_1p_2k_1k_2}\right]_{\mu\nu}&\equiv& \langle p_1,p_2, g_{\mu}|S^0|k_1,k_2,g_{\nu}\rangle =\frac{1}{\sqrt{2}\pi}\int dp_1dp_2\left(e^{-ip_1x_1}e^{-ip_2x_2}+e^{-ip_1x_2}e^{-ip_1x_2}\right)\,\psi_{\text{out}}(x_1,x_2)\nonumber\\
&=&\sum_{P, Q}\sum_{\lambda=1}^2 \frac{i}{2\pi}\,\frac{t_{\mu\lambda}(k_{P(2)}) t_{\lambda\nu} (k_{P(1)})}{p_{Q(2)}-\Delta_{\mu}-k_{P(2)}+\Delta_{\lambda}+i0^+}\delta(p_1+p_2-\Delta_{\mu}-k_1-k_2+\Delta_{\nu})\,,
\end{eqnarray} 
where $P$ and $Q$ are permutation operators that act on indices $1,2$. In (\ref{seq}), the denominator arises from the arguments above regarding sequential scattering. When $\Delta_{\mu}+\Delta_{\nu}=2\Delta_{\lambda}$,
one recovers the familiar form of $S^0$ that contains two $\delta$ functions. Here however, the $S^0$ contains only a single $\delta$ function. Therefore, in the sequential scattering process, the single photon energy is not conserved, 
if the incident wave is the symmetrized plane wave as shown in (\ref{iw}). 

To validate the heuristic arguments above that lead to  (\ref{seq}), we compute the two-photon S matrix $\left[\mathbf{S}_{p_1p_2k_1k_2}\right]_{\mu\nu}\equiv\langle p_1,p_2, g_{\mu}|S|k_1,k_2,g_{\nu}\rangle$ for the Hamiltonian (\ref{int}). 
Using the input-output formalism \cite{fks, gc, xf}, the two-photon S matrix is related to the Green functions of the atom as
\begin{eqnarray}\label{tttt}
\left[\mathbf{S}_{p_1p_2k_1k_2}\right]_{\mu\nu}&=&\frac{1}{2}\delta_{\mu,\nu}\sum_{P,Q}\delta\left(p_{Q(1)}-k_{P(1)}\right)\delta\left(p_{Q(2)}-k_{P(2)}\right)\nonumber\\
&&-\sum_{P,Q}\int \frac{dt'}{\sqrt{2\pi}} e^{ip_{Q(1)} t'}\int \frac{dt}{\sqrt{2\pi}}e^{-ik_{P(1)} t}\langle g_{\mu}|{\cal{T}}A(t')A^{\dag}(t)|g_{\nu}\rangle\,\delta\left(p_{Q(2)}-k_{P(2)}\right)\nonumber\\
&&+\int \frac{dt_1'}{\sqrt{2\pi}} e^{ip_1 t_1'}\int \frac{dt_2'}{\sqrt{2\pi}} e^{ip_2 t_2'}\int \frac{dt_1}{\sqrt{2\pi}}
e^{-ik_1 t_1}\int \frac{dt_2}{\sqrt{2\pi}} e^{-ik_2 t_2}\langle g_{\mu}|{\cal{T}}
A(t'_1)A(t'_2)A^{\dag}(t_1)A^{\dag}(t_2)|g_{\nu}\rangle\,,
\end{eqnarray}
where $A=\sum_{\lambda=1}^2\sqrt{\gamma_{\lambda}}|g_{\lambda}\rangle\langle e|$. The Green functions can be computed by diagonalizing the effective Hamiltonian 
\begin{equation}
H_{\text{eff}}=\sum_{\lambda=1}^2 \widetilde{\Delta}_{\lambda} |g_{\lambda}\rangle\langle g_{\lambda}|+\left(\Omega-i\frac{\gamma_1}{2}-i\frac{\gamma_2}{2} \right)|e\rangle\langle e|
\end{equation}
that is obtained after integrating out the waveguide degrees of freedom. For notation simplicity, we define
\begin{equation}
s_{\mu\nu}(k)\equiv-i\frac{\sqrt{\gamma_{\mu}\gamma_{\nu}}}{k-\Delta_{\nu}+i\left(\frac{\gamma_1}{2}+\frac{\gamma_2}{2}\right)}
\end{equation}
which is related to the transmission amplitude $t_{\mu\nu}(k)$ defined in (\ref{tuv}) as $t_{\mu\nu}(k)=\delta_{\mu\nu}+s_{\mu\nu}(k)$. As a result, we have
\begin{eqnarray}\label{eleS}
\left[\mathbf{S}_{p_1p_2k_1k_2}\right]_{\mu\nu}&=&\frac{1}{2}\delta_{\mu,\nu}\sum_{P,Q}\delta\left(p_{Q(1)}-k_{P(1)}\right)\delta\left(p_{Q(2)}-k_{P(2)}\right)\nonumber\\
&&+\sum_{P,Q}s_{\mu\nu}\left(k_{P(1)}\right)\delta\left(p_{Q(2)}-k_{P(2)}\right)\delta(p_1+p_2-\Delta_{\mu}-k_1-k_2+\Delta_{\nu})\nonumber\\
&&+\sum_{P,Q}\frac{i}{2\pi}\frac{s_{\lambda\mu}\left(p_{Q(2)}\right)s_{\lambda\nu}\left(k_{p(1)}\right)}{p_{Q(2)}-\Delta_{\mu}-k_{Q(2)}+\Delta_{\lambda}+i0^+}\delta(p_1+p_2-\Delta_{\mu}-k_1-k_2+\Delta_{\nu})\,.
\end{eqnarray}
We can further simplify (\ref{eleS}) into the following compact form:
\begin{eqnarray}
\left[\mathbf{S}_{p_1p_2k_1k_2}\right]_{\mu\nu}=\left[\mathbf{S}^0_{p_1p_2k_1k_2}\right]_{\mu\nu}+i\,\left[\mathbf{T}_{p_1p_2k_1k_2}\right]_{\mu\nu}\,,\label{twoS}
\end{eqnarray}
where $\left[\mathbf{S}^0_{p_1p_2k_1k_2}\right]_{\mu\nu}$ is the same as obtained in (\ref{seq})  but now with a rigorous calculation.
$\left[\mathbf{T}_{p_1p_2k_1k_2}\right]$
is the the photon-photon interacting part whose $(\mu, \nu)$ entry is
\begin{eqnarray}
i\,\left[\mathbf{T}_{p_1p_2k_1k_2}\right]_{\mu\nu}&=&-\frac{i}{2\pi}\left(\frac{1}{p_1-\Delta_{\mu}+i\left(\frac{\gamma_1}{2}+\frac{\gamma_2}{2}\right)}
+\frac{1}{p_2-\Delta_{\mu}+i\left(\frac{\gamma_1}{2}+\frac{\gamma_2}{2}\right)}\right)\times\nonumber\\
&&\left[\sum_{\lambda=1}^2\left(s_{\mu\lambda}(k_2)s_{\lambda\nu}(k_1)+s_{\mu\lambda}(k_1)s_{\lambda\nu}(k_2)\right)\right]\delta(p_1+p_2-\Delta_{\mu}-k_1-k_2+\Delta_{\nu})\,.\label{Tmu}
\end{eqnarray}
The interacting part of S matrix as represented by (\ref{Tmu}) now only contains a single $\delta$ function and single-photon excitation poles, which agrees with the cluster decomposition principle \cite{srf}.

With the two-photon S matrix (\ref{twoS}), we now confirm the heuristic argument presented in Fig.\ref{fig1} by an explicit calculation. We
 consider the scattering event of two sequential single
photon pulses spatially well separated from each other. By the identical-particle postulate the two-photon in-state
has the form
\begin{equation}\label{2p}
|\,\bar{k}_1,\bar{k}_2,L, g_{\nu}\rangle\equiv\frac{1}{\sqrt{2}}\left[|\bar{k}_2\rangle\otimes e^{-i\hat{p}L} |\bar{k}_1\rangle+|\bar{k}_1\rangle\otimes
e^{-i\hat{p}L}|\bar{k}_2\rangle\right] \otimes |g_{\nu}\rangle\,,
\end{equation}
where $|\bar{k}\rangle=\int dk\, f_{\bar{k}}(k)\,|k\rangle$ describe
a single photon pulse with mean momentum $\bar{k}$ \footnote{For
example, we could take the envelop of the pulse to be Lorentzian
like $f_{\bar{k}}(k)=\frac{\alpha}{\pi(\alpha^2+(k-\bar{k})^2)}$.}. $\hat{p}$ is the momentum operator and
$L$ is the spatial separation between two pulses. When $L$ is large enough, 
there should be no photon-photon interaction. Indeed,  one can check explicitly 
that (\ref{Tmu}) satisfies the requirement
\begin{equation}\label{LT}
\lim_{L\rightarrow\infty}T\, |\bar{k}_1,\bar{k}_2,L,g_{\nu}\rangle=0\,.
\end{equation}
As a result, the out-state all comes from the non-interacting part of S matrix (\ref{seq}), that is,
\begin{eqnarray}
|\text{out}\rangle&=&\lim_{L\rightarrow\infty}S^0\, |\bar{k}_1,\bar{k}_2,L,g_{\nu}\rangle\nonumber\\
&=& \lim_{L\rightarrow\infty}\frac{1}{4}\sum_{\mu=1}^2\int dp_1dp_2 |p_1,p_2,g_{\mu}\rangle \int dk_1dk_2 \left[\mathbf{S}^0_{p_1p_2k_1k_2}\right]_{\mu\nu}\langle k_1,k_2,g_{\nu}|\,\bar{k}_1,\bar{k}_2,L, g_{\nu}\rangle\,,\nonumber\\
&=&\frac{1}{\sqrt{2}}\sum_{\mu,\lambda=1}^2\left[|\bar{k}_2\rangle_{\mu\lambda}\otimes e^{-i\hat{p}L} |\bar{k}_1\rangle_{\lambda\nu}+|\bar{k}_1\rangle_{\mu\lambda}\otimes
e^{-i\hat{p}L}|\bar{k}_2\rangle_{\lambda\nu}\right] \otimes |g_{\mu}\rangle\,,\label{wa}
\end{eqnarray}
where $|\bar{k}\rangle_{\lambda\nu}\equiv \int dk\, t_{\lambda\nu}(k) f_{\bar{k}-\Delta_{\nu}+\Delta_{\lambda}}(k)|k\rangle $ describes the outgoing single photon pulse with mean momentum $\bar{k}-\Delta_{\nu}+\Delta_{\lambda}$
after scattering.
By comparing the initial state (\ref{2p}) and the final state (\ref{wa}), one can see that
our main result (\ref{seq}) indeed preserves the sequential ordering as represented by the translation operator $e^{-i\hat{p}L}$, and thus produces the correct result of sequential scattering that agrees with previous thought experiment.

The results above can be straightforwardly generalized to other systems supporting multiple ground states, including optomechanical cavities \cite{ll2}
which also contains multiple ground states due to the phonon side bands. Here, by multiple ground states, we include the cases where the ground state manifolds contain metastable states, 
as long as the lifetime of these states significantly exceed the relevant interaction or scattering time-scales \cite{jrtaylor}. 
For a general waveguide QED system consisting of a single mode waveguide coupled to a cavity
\begin{equation}
H=\int dk\,k\,c_{k}^{\dag}c_k+\sqrt{\frac{\gamma}{2\pi}}\int dk\left(c_k^{\dag}a+a^{\dag}c_k\right)+H_c[a,b]\,,
\end{equation}
where $H_c[a,b]$ is the cavity's Hamiltonian. $b$ denotes the other degrees of freedom of the cavity which could be a multi-level atom
or phonons in an optomechanical cavity.  One can integrating out the waveguide photons to obtain an effective Hamiltonian of the cavity \cite{sfs,xf,xfFano}
\begin{equation}\label{effHa}
H_{\text{eff}}[a,b]=H_c[a,b]-i\frac{\gamma}{2} a^{\dag}a\,.
\end{equation}
We also assume that there exits some total excitation operator of the form $\hat{N}=a^{\dag}a+\hat{O}(b)$
such that $\hat{O}\geq 0$ and $\left[\hat{N}, H_{\text{eff}}\right]=0$ \footnote{For example, $\hat{N}=a^{\dag}a+\sigma_z/2$ in the Jaynes-Cummings model and 
$\hat{N}= a^{\dag}a$ in the optomechanical cavity.}. With such $\hat{N}$, $H_{\text{eff}}[a, b]$ can be block diagonalized as
\begin{equation}\label{eev}
H_{\text{eff}}\, |\lambda\rangle_N={\cal{E}}_N^{\lambda}\,|\lambda\rangle_N\,,\,\,\,\,\,\, {}_N\langle\bar{\lambda}|\,H_{\text{eff}}={}_N\langle\bar{\lambda}|\,{\cal{E}}_N^{\lambda}.
\end{equation}
Because $H_{\text{eff}}$ in (\ref{effHa}) is non-Hermitian, its eigenvalues ${\cal{E}}_N^{\lambda}$ are in general complex,
except for a set of ground states $|g_{\lambda}\rangle$ which has zero excitation and hence real eigenvalue $E_0^\lambda$.
 Using the input-output formalism \cite{xf}, we can compute the general
single photon S matrix as
\begin{equation}\label{SSF}
\left[\mathbf{S}_{pk}\right]_{\mu\nu}\equiv \langle p, g_{\mu}|S|k,g_{\nu}\rangle = t_{\mu\nu}(k)\, \delta(p+E_0^{\mu}-k-E_0^{\nu})\,,
\end{equation}
with
\begin{eqnarray}
t_{\mu\nu}(k)=\delta_{\mu\nu}+\sum_{\rho}s_{\nu}^{\rho}(k)\langle g_{\mu}|a|{\rho}\rangle_1 \,{}_1\langle \bar{\rho} |a^{\dag}|g_{\nu}\rangle\,,\,\,\,\,\,\,\,
s_{\nu}^{\rho}(k)\equiv-i\frac{\gamma}{k+E_0^{\nu}-{\cal{E}}_1^{\rho}}\label{suv}\,.
\end{eqnarray}
where we insert the biorthogonal basis as defined in (\ref{eev}) to compute the cavity's Green function \cite{sfs}. 
Using a formula similar to (\ref{tttt}),  the two-photon S matrix can be computed as
\begin{equation}
\mathbf{S}_{p_1p_2k_1k_2}=\mathbf{S}^0_{p_1p_2k_1k_2}+i \mathbf{T}_{p_1p_2k_1k_2}\,,
\end{equation}
where
\begin{eqnarray}
\left[\mathbf{S}^{0}_{p_1p_2k_1k_2}\right]_{\mu\nu}
&=&\sum_{P, Q}\sum_{\lambda} \frac{i}{2\pi}\,\frac{t_{\mu\lambda}(k_{P(2)}) t_{\lambda \nu} (k_{P(1)})}{p_{Q(2)}+E_0^{\mu}-k_{P(2)}-E_0^{\lambda}+i0^+}\delta(p_1+p_2+E_0^{\mu}-k_1-k_2-E_0^{\nu})\,,\label{gseq}\\
i \left[\mathbf{T}_{p_1p_2k_1k_2}\right]_{\mu\nu}&=&i \left[\mathbf{C}_{p_1p_2k_1k_2}\right]_{\mu\nu}\delta(p_1+p_2+E_0^{\mu}-k_1-k_2-E_0^{\nu})\,,\label{gT}\\
i \left[\mathbf{C}_{p_1p_2k_1k_2}\right]_{\mu\nu}&=&
\frac{1}{2\pi\gamma}\sum_{\lambda\rho\sigma}\left[s_{\mu}^{\rho}(p_1)+s_{\mu}^{\rho}(p_2)\right]\left[s_{\lambda}^{\rho}(k_1)s_{\nu}^{\sigma}(k_2)+s_{\lambda}^{\rho}(k_2)s_{\nu}^{\sigma}(k_1)\right]
\langle g_{\mu}| a|{\rho}\rangle_{1}\, {}_1\langle \bar{\rho} |a^{\dag}|g_{\lambda}\rangle \langle g_{\lambda} | a|{\sigma}\rangle_{1} \,{}_1\langle \bar{\sigma} |a^{\dag}|g_{\nu}\rangle
\nonumber\\
&&+\frac{i}{2\pi}\sum_{\lambda\rho\sigma} \left[s_{\mu}^{\rho}(p_1)+s_{\mu}^{\rho}(p_2)\right]\left[s_{\nu}^{\sigma}(k_1)+s_{\nu}^{\sigma}(k_2)\right]
\frac{ \langle g_{\mu}| a|{\rho}\rangle_1\, {}_1\langle \bar{\rho} |a|\lambda\rangle_2\,{}_2\langle\bar{\lambda} | a^{\dag}| {\sigma}\rangle_{1} \,{}_1\langle \bar{\sigma} |a^{\dag}|g_{\nu}\rangle}{k_1+k_2+E_0^{\nu}-{\cal{E}}_2^{\lambda}}\,.
\end{eqnarray}
In the above decomposition, the $T$ matrix (\ref{gT}), which describes the effect of photon-photon interaction, only contains single and two excitation poles as well as a single $\delta$ function related to the energy conservation, as required by the cluster decomposition principle 
 \cite{srf}. The non-interacting part of S matrix (\ref{gseq}) becomes the usual direct product of two single-photon S matrix only in the cases of a single ground state or multiple degenerate ground states.
In general, however, $S^0$ is not a direct product of the single photon S matrix. 

In summary, we present the general structure of two-photon S matrix for a waveguide coupled to a local quantum system with multiple ground states. 
Such two-photon S matrix has an analytic structure that differs significantly from the standard form of the two-particle S matrix in quantum field theory. We show that such a structure arises from a non-commutivity between the two-photon S matrix and an operator that exchanges photon orders. Our results here points to significant additional richness in the analytic structure of S matrix as compared to commonly anticipated. The results also provide a complete description of photon-photon interaction in several waveguide QED systems, including systems with quantum emitters with multiple ground states and systems with optomechanical cavities, that are of importance for on-chip manipulation of photon-photon interactions.

This research is supported by an AFOSR-MURI program, Grant No. FA9550-12-1-0488.


\end{document}